\documentclass{iopartbfsubs}

\usepackage{epsfig}
\usepackage{times}

\usepackage{color}
\definecolor{dkgreen}{rgb}{0,.3,0}
\definecolor{dkred}{rgb}{.7,.0,.0}

\newcommand{\cwg}[1]{#1}
\newcommand{\mjd}[1]{#1}
\renewcommand{\note}[1]{}

\newcommand{\Label}[1]{\label{#1}}

\newcommand{\GPE}{Gross-Pitaevskii equation}
\newcommand{\FTGPE}{finite temperature Gross-Pitaevskii equation}

\newcommand{\BEC}{Bose-Einstein condensate}
\newcommand{\BECn}{Bose-Einstein condensation}
\newcommand{\SDE}{stochastic differential equation}
\newcommand{\FPE}{Fokker-Planck equation}
\newcommand{\QKIII}{\cite{Gardiner1998a}}

\newcommand{\grapprox}{\mathrel{\vbox{\hbox{$\sim$}\vskip -7.5pt\hbox{$>$}
\vskip-1pt}}}
\newcommand{\ltapprox}{\mathrel{\vbox{\hbox{$\sim$}\vskip 
-7.5pt\hbox{$<$}
\vskip-1pt}}}
\newcommand{\deltaC}{\bar\delta}
\newcommand{\LC}{\bar L_{C}}

\newcommand{\Sect}{Sect.\,}
\newcommand{\Sects}{Sects.\,}
\renewcommand{\etal}{{\em et al.}}  
\newcommand{\Ppt}[1]{\vskip 1pt\noindent{#1) }}

\newcommand{\DRAFT}{%
\renewcommand{\cwg}[1]{{\color{dkgreen}##1}}%
\renewcommand{\mjd}[1]{{\color{dkred}##1}}%
\renewcommand{\note}[1]{\par\noindent{\color{blue}\emph{##1}\par\noindent}}%
\renewcommand{\Label}[1]{\label{##1}{\hbox to 0cm{%
\textcolor{magenta}{\hss\em ##1\quad}}}}}

\usepackage{graphicx}
\usepackage{color}

\renewcommand{\include}[1]{\input{#1.tex}}

\begin{document}
\title{The stochastic \GPE: II}
\author{C. W. Gardiner$^{1}$ and M. J. Davis$^{2}$}
\address{$^{1}$School of Chemical and Physical Sciences, Victoria University 
of Wellington, New Zealand.}
\address{$^{2}$ARC Centre of Excellence for Quantum Atom Optics, Department of
Physics, University of Queensland, St Lucia, QLD 4072, Australia.}

\begin{abstract}
We provide a derivation of a more accurate version of the stochastic
\GPE, as introduced by Gardiner \etal~\cite{Gardiner2002a}. The
derivation does not rely on the concept of local energy and
momentum conservation, and is based on a quasi-classical
Wigner function representation of a ``high temperature'' master
equation for a Bose gas, which includes only modes below an energy
cutoff $E_R$ that are sufficiently highly occupied (the condensate
band). The modes above this cutoff (the non-condensate band) are
treated as being essentially thermalized. The interaction between
these two bands, known as growth and scattering processes, provide
noise and damping terms in the equation of motion for the condensate
band, which we call the stochastic \GPE. This approach is distinguished by
the control of the approximations made in its derivation, and by the
feasibility of its numerical implementation.
\end{abstract}

\section{Introduction}\Label{introduction}
Notwithstanding the large body of work done on the kinetics and dynamics of 
\BEC s \cite{kinetics}, there is still a need for a method of treating these 
aspects which is 
both accurately related to fundamental theory and at the same time able to be 
implemented practically and reliably.  Problems for which this would be 
particularly useful are those of condensate growth, nucleation of vortices and 
the treatment of heating by mechanical disturbance.  

In this paper we will 
develop the theoretical basis for a description based on a \SDE\ for a 
quasiclassical field, which arises from a Wigner function representation 
of the quantum field. Descriptions of this kind have been presented previously 
by Stoof \cite{Stoof1999a} and ourselves \cite{Gardiner2002a}.  Related phase 
space methods have been presented, which use the Wigner function either 
explicitly \cite{Steel1998a,Sinatra2000a,Sinatra2001a,Sinatra2002a}
or implicitly 
\cite{Davis2001a,Davis2001b,Davis2002a,Goral2001a,Goral2002a,Schmidt2003a} with 
random initial 
conditions to simulate  quantum 
noise have also been presented, and all have shown that a significant 
proportion of experimental reality can be reproduced.

The method used here can be seen as a unification of 
the ideas of \emph{quantum kinetic theory} as presented in 
\cite{Gardiner1997a,Gardiner1998a,Gardiner2000a}
with those of the \emph{\FTGPE}, as developed by Davis and co-workers
\cite{Davis2001a,Davis2001b,Davis2002a}. The main idea is that 
the higher energy modes of a Bose gas are largely thermalized and can be 
eliminated, to produce a quantum mechanical master equation. This is also 
the central concept in Stoof's work---we compare our methodology with his
in \Sect{\ref{Stoof}}.  

We do the elimination in two stages:
\begin{itemize}
\item [a)] We first eliminate modes with a wavenumber $ {\bf k}$ such that
$ |{\bf k}| >\Delta$, where $2\pi/ \Delta $ is of order of magnitude of the 
range of the interatomic potential.  Under conditions normally met in \BEC s, 
these modes have no occupation, and the effect is to remove from consideration 
the high momentum components which occur during an actual collision.  This 
leads to a ``coarse grained'' quantum field theory which contains no high 
momentum components, and which can quite accurately be described by a simple 
\nocite{Fermi1936a,Fermi1936b}
Fermi delta function pseudopotential \cite{Fermi1936a,Breit1947a,Blatt1952a}.  
For this kind of elimination there is no 
need to use the many-body T-matrix formulation, as there would be if $ \Delta$ 
were much smaller, and we were required to eliminate  thermally occupied 
states; neither is it necessary to introduce the Huang-Yang pseudopotential
\cite{Huang1957a} involving the derivative of a delta function.

\item [b)]  We then separate the remaining momentum range into a low momentum 
component \mjd{(the condensate band)}, and a high momentum component 
\mjd{(the  non-condensate band)}. We treat the condensate band fully
quantum-mechanically,  while the non-condensate band is treated as a bath of
thermalized atoms, which  in this paper and \cite{Gardiner1998a} is considered
to be unchanging or in  \cite{Gardiner2000a}  is treated by a kind of quantum
Boltzmann equation. 

\item [c)] The condensate band contains much more than simply the condensate.  
Typically it spans an energy range of the order of magnitude of twice the 
chemical potential.  In this paper, our criterion will be that all modes in the 
condensate band should be have sufficient occupation (which we take to be more 
than 10 atoms) for us to be able to make 
the approximations necessary for a Wigner function \SDE\ to be valid. 
\item [d)] The 
resulting \SDE\ is very similar to Stoof's in general appearance, but has 
extra 
noise terms, and no reference to the many-body T-matrix or self energy 
functions.  These effects are provided by solving the \SDE\ itself. 
It also explicitly contains the projector into the condensate band, which 
ensures that the solutions of the \SDE\ remain within the condensate band.
\end{itemize}

\noindent
The organization of the paper is as follows.
We outline our description of the system in \Sect\ref{Desc} and
present the derivation of the corresponding 
master equation in \Sect\ref{Master}.   We then consider two simplifications
of the master equation in \Sect\ref{Approx} which we call:
\begin{itemize}
\item [a)] The ``quantum optical'' master
 equation, which corresponds to the methodology of our earlier Quantum Kinetic 
Theory papers \cite{Gardiner1997a,Gardiner1998a,Gardiner2000a};
\item [b)] The ``high temperature'' master equation, whose validity requires
$ k_BT $ to be significantly larger than the particle or quasiparticle energies 
described.
\end{itemize}
In \Sect\ref{Wigner} a
Fokker-Planck equation is derived from
the latter form of the master equation, and this is equivalent to a set of 
\SDE s that correspond to the  stochastic \GPE\ of the title.  
These \SDE s are very similar to those proposed by Stoof \cite{Stoof1999a} and 
ourselves \cite{Gardiner2002a}, but differ in kind of noise considered, and in 
the explicit implementation of the projection techniques of Davis \etal\ 
\cite{Davis2001a,Davis2001b,Davis2002a}.
Finally we conclude in \Sect\ref{conclusions}, with a discussion of the range 
of applicability of the methodology developed, and suggest systems \mjd{that}
should be investigated \mjd{within this framework.}

\section{Description of the system}\Label{Desc}
\subsection{The cold-collision Hamiltonian}\Label{cold-collision}
A system of Bose atoms interacting via an interatomic potential 
$ u({\bf x})$ is almost universally simplified in order to separate the short 
distance dynamics of pairs of atoms as they proceed through scattering events 
from what one normally considers to be the interesting collective behaviour of 
the gas itself, which takes place on a relatively slow time scale and over 
larger distance scales.  There are a number of ways in which one can proceed to 
an 
appropriately ``coarse grained'' description, in which the details of the 
interaction are replaced by a single parameter $ a$, the scattering length for 
interatomic collisions.  All methods, either implicitly or explicitly, 
introduce a momentum scale $ \hbar\Delta$, above which all degrees of freedom 
are eliminated.  For the description in terms of the scattering length to be 
valid, there must be essentially no occupation of the states eliminated, and 
this requires
\begin{eqnarray}\Label{2001}
k_BT &\ll& {\hbar^2\Delta^2/ 2m}.
\end{eqnarray}
If $ \Delta$ does not satisfy this criterion, a description in terms of the 
\emph{many-body T-matrix} rather than simply the scattering length is required, 
as is done in the approach of Stoof \cite{Stoof1999a}.

The criterion that motion of the particles inside the range of the interatomic 
potential be eliminated leads to the requirement 
\begin{eqnarray}\Label{2002}
\Delta &\ll  & {2\pi/ r_0},
\end{eqnarray}
where $ r_0$ is the effective range of the interatomic potential.  These two 
conditions together obviously lead to a condition on the temperature
\begin{eqnarray}\Label{2003}
k_BT &\ll& {h^2/ 2mr_0^2}.
\end{eqnarray}
Under the further condition that
\begin{eqnarray}\Label{2004}
a\Delta&\ll& 1,
\end{eqnarray}
we can describe the dynamics of the modes with momenta below the cutoff 
$ \hbar\Delta$ by a field operator $ \psi({\bf x})$, which contains modes only 
with momentum below the cutoff, and a Hamiltonian
\begin{eqnarray}\Label{Desc1.1}
H=H_{\rm sp}+H_I ,
\end{eqnarray}
in which the \emph{single particle} Hamiltonian is
\begin{eqnarray}\Label{Desc1.2}
H_{\rm sp}=\int d^3{\bf x}\,\psi ^{\dagger }({\bf x})
\left(- {\hbar ^2\over2m}\nabla ^2 +V({\bf x})\right) \psi ({\bf x})  , 
\end{eqnarray}
and the \emph{interaction Hamiltonian}
\begin{eqnarray}\Label{Desc1.3}
H_I&=&{u\over 2}\int d^3{\bf x}\,\psi ^{\dagger }({\bf x})
\psi ^{\dagger }( {\bf x})
\psi ( {\bf x}) \psi ( {\bf x}) . 
\end{eqnarray}
Effectively, the elimination of the modes above the cutoff has made the 
replacement to the interatomic potential 
$ u({\bf x})\to u\delta({\bf x})$, with
\begin{eqnarray}\Label{2005}
u &=& 4\pi\hbar^2 a/m ,
\end{eqnarray}
and, it must be emphasized, the resulting field theory has the cutoff, 
$ \Delta$, so that the commutation relations of the field operator are
\begin{eqnarray}\Label{2006}
\left[\psi({\bf x}),\psi^\dagger({\bf x}')\right]
 &=& {\cal P}_\Delta({\bf x}-{\bf x}').
\end{eqnarray}
The function  ${\cal P}_\Delta $ plays the r\^ole of a kind of coarse-grained 
delta function; however, it is also a projector into the subspace of 
non-eliminated modes.  Using the commutation relation (\ref{2006})
the Heisenberg equation of motion for the field operator takes the form
\begin{eqnarray}\fl\Label{2007}
\rmi\hbar{\partial\psi({\bf x})\over\partial t}&=&
\int d^3{\bf x}'\,{\cal P}_\Delta({\bf x}-{\bf x}')
\left\{-{\hbar^2\over 2m}\nabla^2\psi({\bf x}') +V({\bf x}')
 + u \psi^\dagger({\bf x}')\psi({\bf x}')\psi({\bf x}')\right\}.
\end{eqnarray}
In practical computations a momentum cutoff is selected by the spatial grid 
used, and on this scale $ {\cal P}_\Delta$ appears like a delta function.

When all of the conditions (\ref{2001}--\ref{2004}) are satisfied, 
this Hamiltonian is well-defined, and the Born series is both 
convergent and accurate at first order.

\subsection{Condensate and non-condensate bands}\Label{Bands}
We now divide the states of the system
into a {\em condensate band } $ R_C$ and {\em non-condensate band} $ R_{NC}$, 
and  perform a corresponding resolution of the field operator in the form
\begin{eqnarray}\Label{1999}
 \psi({\bf x})& = &\phi({\bf x}) + \psi_{NC}({\bf x}).
\end{eqnarray}
We will describe $ 
R_C$ fully quantum-mechanically, while $ R_{NC}$ will be taken as being 
essentially thermalized.
The cut between $ R_C$ and $ R_{NC}$ is set in terms of the  single particle 
energy $ E_R$, which is such that particles with higher energy than this can be 
considered to be fully thermalized with very little effect on their energies 
from the condensate band.

We want to make clear at this stage that $ E_R \ll \hbar^2\Delta^2/2m$,  and
thus that the division into condensate and non-condensate bands is quite 
independent  of the cut at the wavenumber $ \Delta$ treated in the previous
section.  
\mjd{This is essentially the procedure followed in our work on the 
\FTGPE~\cite{Davis2001a,Davis2001b,Davis2002a} with the use of the contact
potential approximation $U_0 \delta(\mathbf{x})$ in the spatial representation
of the equations of motion.  However, in \cite{Davis2001a} the FTGPE in basis
notation is written in terms of the the full interatomic potential, and 
\Sect{5.2} of \cite{Davis2001a} illustrates how certain terms can upgraded to a 
T-matrix description that is well approximated by a contact potential.
The point of view adopted in this paper is consistent 
with our use of the delta function potential in quantum kinetic theory 
\cite{Gardiner1997a,Gardiner1998a,Gardiner2000a}.
}

\begin{figure}[t]
\epsfig{file=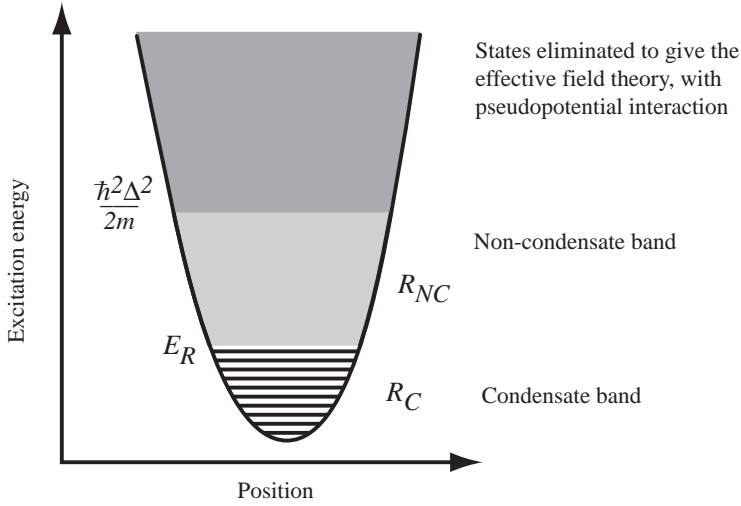,height=6.75cm}
\caption{\Label{fig1.eps}%
Schematic view of the condensate band, the non-condensate band, and eliminated 
states for  a harmonic trap.}
\end{figure}

For the purposes of this paper we have to take into account two criteria:
\begin{itemize}
\item [a)] The highest energy states of the condensate band should have an 
occupation of the order of magnitude 5--10, so that we can apply Wigner 
function methods to \emph{all} modes of the condensate band.  In practice this 
means that
\begin{eqnarray}\Label{2008}
{k_BT\over E_R-\mu} &\grapprox & 5,
\end{eqnarray}
since this corresponds to the high  occupation limit of the Bose-Einstein 
distribution.
\item [b)] The single particle energy levels of the  noncondensate band should 
be essentially those of the trapping potential $ V({\bf x})$, which is usually 
a harmonic potential.  In \cite{Gardiner1998a} we showed that this is true to 
about 3\%, provided that 
\begin{eqnarray}\Label{2009}
E_R/\mu_C & \grapprox & 3.
\end{eqnarray}
The two criteria together lead to the requirement (assuming that the 
maximum value of $ \mu_C\approx 
\mu$) that
\begin{eqnarray}\Label{2010}
k_BT/\mu &\grapprox & 10,
\end{eqnarray}
and this condition is usually satisfied in practice.
\end{itemize}

\subsubsection{Definition of condensate band field operators}
To effect the division between the two bands in the many-particle Hamiltonian
we first note that the effective potential in the noncondensate band is 
considered to be little different from the trap potential.  Thus
we can expand $ \psi({\bf x})$ in trap eigenfunctions, and define 
$ \psi_{NC}({\bf x}) $ to be that component which is expressible entirely in 
terms of eigenfunctions with energy eigenvalue larger than $ E_R$.
This means that 
 $ \phi({\bf x})$ is automatically defined through (\ref{1999}), and that
it can be expressed in terms of  trap eigenfunctions with energy eigenvalue 
less than $ E_R$.

The division can be expressed in terms of projectors, whose form will be 
important in numerical implementations, as
\begin{eqnarray}\Label{1001}
{\cal P}_{NC}({\bf x},{\bf x}') &\equiv&
 \sum_{E_n>E_R}Y_n^*({\bf x})Y_n({\bf x}'),
\\ \Label{1002}
{\cal P}_{C}({\bf x},{\bf x}') &\equiv& 1 -{\cal P}_{NC}({\bf x},{\bf x}') .
\end{eqnarray}
Here the $ Y_n({\bf x})$ are a complete set of trap eigenfunctions.  Thus
\begin{eqnarray}\Label{1003}
\psi_{NC}({\bf x}) &=& 
\int d^3{\bf x}'\, {\cal P}_{NC}({\bf x},{\bf x}')\psi({\bf x}')
\equiv {\cal P}_{NC}\big\{\psi({\bf x})\big\},
\\ \Label{1004}
\phi({\bf x}) &=& 
\int d^3{\bf x}'\, {\cal P}_{C}({\bf x},{\bf x}')\psi({\bf x}')\,\,\,\,
\equiv {\cal P}_{C}\big\{\psi({\bf x})\big\}.
\end{eqnarray}

\subsubsection{Separation of condensate and non-condensate parts of the full 
Hamiltonian}
The Hamiltonian is written in a form which separates it into three 
components; namely, those which act within $ R_{NC}$ only, those which act 
within $ R_C $ only, and those which cause transfers of energy or population 
between $ R_C$ and $ R_{NC}$.  Thus we write
\begin{eqnarray}\Label{H1}
H =  H_{NC} + H_{0} +H_{I,C},
\end{eqnarray}
in which $ H_{NC}$ is the part of $ H$ which depends only on $ \psi_{NC}$, 
$ H_0$ is the part which depends only on $ \phi$, and that part of the 
interaction Hamiltonian which involves both condensate band and non-condensate 
band operators
is called $ H_{I,C}$. Substituting $ \psi({\bf x})\to \phi({\bf x}) + 
\psi_{NC}({\bf x})$ in the Hamiltonian, we get
\begin{eqnarray}\Label{HIC}
H_{I,C}\equiv H^{(1)}_{I,C}+H^{(2)}_{I,C}+H^{(3)}_{I,C},
\end{eqnarray}
where the individual terms in $ H_{I,C}$ are the terms involving operators from 
both bands, which cause 
transfer of energy and/or particles between $ R_C$ and $ R_{NC}$. We call the 
parts involving one $ \phi$ operator
\begin{eqnarray}\Label{HIC1}
H^{(1)}_{I,C}&\equiv &
u\int d^3{\bf x}\,
\left(\psi_{NC}^\dagger({\bf x})\psi_{NC}^\dagger({\bf x})\psi_{NC}({\bf x})
\right)
\phi({\bf x}) 
\nonumber\\ && 
+ u\int d^3{\bf x}\,
\phi^\dagger({\bf x})\left(\psi_{NC}^\dagger({\bf x})\psi_{NC}({\bf x})
\psi_{NC}({\bf x})\right).
\end{eqnarray}
The parts involving  two $ \phi $ operators are called
\begin{eqnarray}\Label{HIC2}
H^{(2)}_{I,C}&\equiv &
2u\int d^3{\bf x}\,
\psi_{NC}^\dagger({\bf x})\psi_{NC}({\bf x})
\Big(\phi^\dagger({\bf x})\phi({\bf x})\Big)
\nonumber\\
&&+{u\over 2} \int d^3{\bf x}\,
\psi_{NC}^\dagger({\bf x})\psi_{NC}^\dagger({\bf x})
\Big(\phi({\bf x})\phi({\bf x})\Big)
\nonumber\\
&&+ {u\over 2}\int d^3{\bf x}\,
\psi_{NC}({\bf x})\psi_{NC}({\bf x})\Big(\phi^\dagger({\bf x})\phi^\dagger({
\bf 
x})\Big).
\end{eqnarray}
The parts involving  three $ \phi $ operators are called
\begin{eqnarray}\Label{HIC3}
 H^{(3)}_{I,C}&\equiv& 
u \int d^3{\bf x}\,
\Big(\phi^\dagger({\bf x})\phi^\dagger({\bf x})\phi({\bf x})\Big)
\psi_{NC}({\bf x}) 
 \nonumber\\  && +u \int d^3{\bf x}\,
\psi_{NC}^\dagger({\bf x})\Big(\phi^\dagger({\bf x})\phi({\bf x})
\phi({\bf x})\Big).
\end{eqnarray}

\subsubsection{Connection to the notation of Davis \it{et al.}}
Here we briefly make a link between the notation of this paper,
which corresponds largely to that of the Quantum Kinetic Theory 
papers~\cite{Gardiner1997a,Gardiner1998a,Gardiner2000a},
and that of the  \FTGPE\ 
papers~\cite{Davis2001a,Davis2001b,Davis2002a}.  
\begin{itemize}
\item [i)] The condensate band $R_{NC}$ 
of Quantum Kinetic Theory roughly corresponds to the coherent region $C$ 
of the  \FTGPE\  papers, although there is the additional requirement in 
the latter that the mode occupations must be large. However, this is also a 
necessary condition for the condensate band in this paper. 
\item [ii)] The non-condensate band $R_{NC}$ was called the incoherent region 
$I$ in \FTGPE\ papers. 
\item [iii)] For the field
operators we have the correspondences
\begin{eqnarray}\Label{corres1}
\phi({\bf x}) \leftrightarrow \hat{\psi}(\mathbf{x}),
\qquad  \psi_{NC}({\bf x}) \leftrightarrow \hat{\eta}(\mathbf{x}).
\end{eqnarray}
\item [iv)] The notations for the projectors are connected by
\begin{eqnarray}\Label{corres2}
{\cal P}_{C}({\bf x},{\bf x}') \leftrightarrow \hat{\mathcal{P}},
\qquad
{\cal P}_{NC}({\bf x},{\bf x}') \leftrightarrow \hat{\mathcal{Q}}.
\end{eqnarray}
\end{itemize}

\section{Derivation of the master equation}\Label{Master}
The  description  assumes that 
$ R_{NC}$ is at least locally thermalized, and 
thus the atoms in these levels are treated simply as a heat bath and 
source of atoms for the levels in $ R_C$. This is defined by requiring the 
field operator correlation 
functions to have a thermal form locally, and to have factorization properties 
like those which pertain in equilibrium.  The precise 
nature of these local equilibrium requirements is specified in
\Sect\ref{Correlation}

\subsection{Formal derivation of the master equation}
The derivation of the master equation follows a rather standard methodology, 
formulated in \cite{QN}, and as previously introduced in 
\cite{Gardiner1997a,Gardiner1998a,Gardiner2000a}.
We  project out the dependence on the 
non-condensate band by defining the condensate density operator as
\begin{eqnarray}\Label{proj1}
\rho_C &=& {\rm Tr}_{NC}(\rho),
\end{eqnarray}
and a projector $ {\cal P}$ on the space of density operators  by
\begin{eqnarray}\Label{proj2}
{\cal P}\rho & = &\rho_{NC}\otimes{\rm Tr}_{NC}(\rho)
\equiv \rho_{NC}\otimes\rho_C ,
\end{eqnarray}
and we also use the notation
\begin{eqnarray}\Label{defQ}
{\cal Q}\equiv 1-{\cal P}.
\end{eqnarray}
The equation of motion for the full density operator $ \rho$ is the von 
Neumann 
equation
\begin{eqnarray}\Label{H8}
\dot \rho &=& -{i\over \hbar}[H_{NC} + H_0 +H_{I}, \rho],
\\ &\equiv &  ({\cal L}_{NC}+{\cal L}_{0}+{\cal L}_{I} ) \rho .
\end{eqnarray}
We use the Laplace transform notation for any 
function $ f(t)$ 
\begin{eqnarray}\Label{Laplace}
\tilde f(s) &=& \int_0^\infty \rme^{-st}f(t)\,dt.
\end{eqnarray}
We then use standard methods to write the master equation for the Laplace 
transform of
$v(t)\equiv {\cal P}\rho(t)$ as
\begin{eqnarray}\fl\Label{H9}
 s \tilde v(s) -  v(0) &=&
{\cal L}_{0}\tilde v(s) 
+ {\cal P}{\cal L}_{I}\tilde v(s)
+{\cal P}{\cal L}_{I}[s- {\cal L}_{NC}-{\cal L}_{0} 
-{\cal Q}{\cal L}_{I}]^{-1}
{\cal Q}{\cal L}_{I}\tilde v(s).
\end{eqnarray}
In this form the master equation is basically exact.  We shall make the 
approximation that the kernel of the second part, the $ [\quad]^{-1}$ term, can
be approximated by keeping only the terms which describe the basic 
Hamiltonians within $ R_C$ or $ R_{NC}$, namely the terms $ {\cal L}_{NC} $ 
and 
$ {\cal L}_{0}$.  We then invert the Laplace transform, and make a Markov 
approximation to get
\begin{eqnarray}\fl\Label{met}
\dot v(t)& =& ({\cal L}_0+{\cal P}{\cal L}_{I})v(t) +
\left\{{\cal P}{\cal L}_{I}\int_0^td\tau\,
\exp\left\{({\cal L}_0+{\cal L}_{I})\tau\right\}
{\cal Q}{\cal L}_{I}\right\} v(t).
\end{eqnarray}
There are now a number of different terms to consider.

\subsection{Terms in the master equation}
\subsubsection{Hamiltonian and forward scattering terms}
These arise from the term 
$({\cal L}_0+{\cal P}{\cal L}_{I})v(t) $,  
and lead to a Hamiltonian term of the form
\begin{eqnarray}\Label{H10}
H_C &\equiv & H_0 + H_{\rm forward},
\end{eqnarray}
where the forward scattering term is defined by
\begin{eqnarray}\Label{H11}
H_{\rm forward}&\equiv&
2u\!\int\! d^3{\bf x} \,\bar n_{NC}({\bf x})
\phi^\dagger({\bf x})\phi({\bf x}),
\end{eqnarray}
where the non-condensate band particle density is
\begin{eqnarray}\Label{H1100}
\bar n_{NC}({\bf x}) &\equiv& {\rm Tr}_{NC}\left\{\psi^\dagger_{NC}({\bf x)}
\psi_{NC}({\bf x})\rho_{NC}\right\}.
 \end{eqnarray}
This represents the condensate band Hamiltonian corrected for the effect of the 
average non-condensate density on the condensate.  This correction is usually 
small, but can be included explicitly in our formulation of the master 
equation.  
The corresponding master equation term is 
\begin{eqnarray}\Label{H1101}
\left. \dot \rho_C\right |_{\rm Ham} &\equiv& -{\rmi\over\hbar}\left[H_C,\rho_C
\right].
\end{eqnarray}

\subsubsection{Interaction between $ R_C$  and $ R_{NC}$}
We now examine the terms in $ H^{(1)}_{I}$, as defined in
(\ref{HIC1}), which contain one
$ \phi({\bf x})$ or $ \phi^\dagger({\bf x})$; explicitly, the term in 
Hamiltonian can be written
\begin{eqnarray}\Label{bba701}
H^{(1)}_{I}&=&
\int d^3{\bf x}\,Z_{3}({\bf x})\phi^\dagger({\bf x})
+ {\rm h.c.}
\end{eqnarray}
In this equation we have defined a notation
\begin{eqnarray}\Label{bba802}
Z_{3}({\bf x}) 
&=& u \psi^\dagger_{NC} ({\bf x})\psi_{NC} ({\bf x})\psi_{NC} ({\bf x}).
\end{eqnarray}
Substituting into the master equation (\ref{met}), terms arise of which a 
typical one is of the form
\begin{eqnarray}\fl\Label{bba8}
&&-{1\over\hbar^2}\int d^3{\bf x}\int d^3{\bf x'}\int_{-\infty}^{\,0} d\tau\,
{\rm Tr}\left\{Z_{3}({\bf x})Z^\dagger_{3}({\bf x}',\tau)\rho_{NC}\right\}
\phi^\dagger({\bf x}) \phi({\bf x'},\tau)
\rho_C(t).
\end{eqnarray}
Here we have introduced the notation for an arbitrary non-condensate band 
operator
\begin{eqnarray}\Label{notationsNC}
A_{NC}({\bf x},t)& =& 
\rme^{\rmi H_{NC}t/\hbar}A_{NC}({\bf x}) \rme^{-\rmi H_{NC}t/\hbar},
\end{eqnarray}
and for an arbitrary condensate band operator
\begin{eqnarray}\Label{notationsC}
B_C({\bf x},t)& =&
 \rme^{\rmi H_{C}t/\hbar}B_C({\bf x}) \rme^{-\rmi H_{C}t/\hbar}.
\end{eqnarray}
This notation will be used frequently in the remainder of this section.

\subsubsection{\Label{Correlation}Correlation functions of $ R_{NC}$}
The terms involving $ Z_3, Z_3^\dagger$ are averaged over $ \rho_{NC}$, which 
is assumed thermalized, and is therefore quantum Gaussian.
This means that:
\Ppt{i} We may make the replacement
\begin{eqnarray}\fl\Label{GaussFactor}
\langle Z_3({\bf x}) Z^\dagger_3({\bf x}',\tau)\rangle
&\to & 2\langle\psi_{NC}({\bf x})\psi^\dagger_{NC}({\bf x}',\tau)\rangle
\langle\psi_{NC}({\bf x})\psi^\dagger_{NC}({\bf x}',\tau)\rangle
\langle\psi^\dagger_{NC}({\bf x})\psi_{NC}({\bf x}',\tau)\rangle  .
\nonumber \\ \fl
\end{eqnarray}
(Terms involving 
$\langle\psi^\dagger_{NC}({\bf x})\psi_{NC}({\bf x}')\rangle$ etc. do arise 
in principle, but give no contribution to the final result because of energy 
conservation considerations.)

\Ppt{ii} The time-dependence is needed only for small $ \tau$, and in this 
case we 
can make  appropriate replacements in terms of the one-particle Wigner 
function $ F({\bf u},{\bf K})$
\begin{eqnarray}\fl\Label{Wig1}
\left\langle\psi^\dagger_{NC}\left({\bf u}+{{\bf v}\over2}\right)
\psi_{NC}\left({\bf u}-{{\bf v}\over2},\tau\right)\right\rangle    
&\approx& {1\over (2\pi)^3}\int_{R_{NC}} d^3{\bf K}\, F({\bf u},{\bf K})
\rme^{-\rmi{\bf K}\cdot{\bf v}-\rmi\omega({\bf K},{\bf u})\tau},
\nonumber\\
\\ \fl
\Label{Wig2}
\left\langle\psi_{NC}\left({\bf u}+{{\bf v}\over2}\right)
\psi^\dagger_{NC}\left({\bf u}-{{\bf v}\over2},\tau\right)\right\rangle
&\approx &{1\over (2\pi)^3}\int_{R_{NC}} d^3{\bf K}\,
\left[ F({\bf u},{\bf K})+1\right]
\rme^{\rmi{\bf K}\cdot{\bf v}+\rmi\omega({\bf K},{\bf u})\tau} ,
\nonumber \\
\end{eqnarray}
where
\begin{eqnarray}\Label{s4}
{\bf u}&\equiv& {{\bf x} + {\bf x}'\over 2},
\\ \Label{s401a}
{\bf v}&\equiv &{\bf x} - {{\bf x}'},
\\ \Label{omegaku}
\hbar\omega({\bf K},{\bf u})&=& {\hbar^2{\bf K}^2\over 2m}+V({\bf u}).
\end{eqnarray}
Since the range of all the $ {\bf K}$ integrals is restricted to $ R_{NC}$, 
it is implicit that in all integrals 
$\hbar\omega({\bf K},{\bf u})>E_R $.

\Ppt{iii} This approximation is valid in the situation where
 $ F({\bf K},{\bf u})$ is a smooth function of its arguments, and can be 
regarded as a local equilibrium assumption for particles moving in a potential
which is comparatively slowly varying in space.

\subsection{Growth terms}
Using (\ref{GaussFactor},\ref{Wig1},\ref{Wig2}) we can now write
\begin{eqnarray}\fl\Label{s2}
\langle Z_3({\bf x}) Z^\dagger_3({\bf x}',\tau)\rangle
&=&
{2u^2\over(2\pi)^9}
\int\!\!\!\!\int\!\!\!\!\int\! d^3{\bf K}_1d^3{\bf K}_2 d^3{\bf  K}_3
[1+F({\bf u},{\bf K}_1)]
[1+F({\bf u},{\bf K}_2)]
F({\bf u},{\bf K}_3)
\nonumber\\ \fl&&
\qquad\qquad
\qquad\qquad\times\rme^{-\rmi(\omega_1+\omega_2-\omega_3)\tau
-\rmi({\bf K}_1+{\bf K}_2-{\bf K}_3)\cdot{\bf v}},
\end{eqnarray}
where
\begin{eqnarray}\Label{s401b}
\hbar\omega_i &\equiv & \hbar\omega({\bf K}_i,{\bf u}).
\end{eqnarray}

\subsubsection{Master equation terms for growth}
We can write 
\begin{eqnarray}\Label{s6}
\phi({\bf x},\tau) &=& \exp(-\rmi L_{C} \tau)\phi({\bf x}),
\end{eqnarray}
with
\begin{eqnarray}\Label{s7}
L_{C} \phi({\bf x}) &\equiv& -[H_{C},\phi({\bf x})]/\hbar .
\end{eqnarray}
This means that (\ref{bba8}) can be written as
\begin{eqnarray}\Label{s8}
\int d^3{\bf u}\int d^3{\bf v}\,\phi^\dagger\left({\bf u}+{{\bf v}\over 2}
\right)
\left\{  G^{(-)}({\bf u},{\bf v}, L_{C})
\phi\left({\bf u}-{{\bf v}\over 2}\right)
\right\}\rho  ,
\end{eqnarray}
with
\begin{eqnarray}\fl\Label{s9}
  G^{(-)}({\bf u},{\bf v}, \omega)
&=&-{2u^2\over(2\pi)^9\hbar^2}\!\!\int\!\!\!\!\int\!\!\!\!\int\!
 d^3{\bf K}_1d^3{\bf K}_2 d^3{\bf K}_3
[1+F({\bf u},{\bf K}_1)]
[1+F({\bf u},{\bf K}_2)]
F({\bf u},{\bf K}_3)
\nonumber\\ \fl&&
\qquad\qquad\times\int_{-\infty}^{\,0} d\tau\,\rme^{\rmi(\omega_1+\omega_2-
\omega_3-
\omega)\tau
-\rmi({\bf K}_1+{\bf K}_2-{\bf K}_3)\cdot{\bf v}} ,
\\ \fl\Label{s901}
&\approx&-{u^2\over(2\pi)^8\hbar^2}\!\!\int\!\!\!\!\int\!\!\!\!\int\!
 d^3{\bf K}_1d^3{\bf K}_2 d^3{\bf K}_3
[1+F({\bf u},{\bf K}_1)]
[1+F({\bf u},{\bf K}_2)]
F({\bf u},{\bf K}_3)
\nonumber\\ \fl&&
\qquad\qquad\times
\delta(\omega_1+\omega_2-\omega_3-\omega)
\rme^{-\rmi({\bf K}_1+{\bf K}_2-{\bf K}_3)\cdot{\bf v}}  .
\end{eqnarray}
(The approximation made in (\ref{s901})  is to say
$ \int_{-\infty}^{\,0} d\tau\exp(-\rmi\Omega\tau)
= \pi\delta(\Omega) + \rmi{\rm P}/\Omega
 \approx \pi\delta(\Omega)$, i.e., to neglect the principal value integral).

We will also need
\begin{eqnarray}\fl\Label{s90101}
  G^{(+)}({\bf u},{\bf v}, \omega)
&\approx&-{u^2\over(2\pi)^8\hbar^2}\!\!\int\!\!\!\!\int\!\!\!\!\int\!
 d^3{\bf K}_1d^3{\bf K}_2 d^3{\bf K}_3
F({\bf u},{\bf K}_1)
F({\bf u},{\bf K}_2)
[1+F({\bf u},{\bf K}_3)]
\nonumber\\ \fl &&
\qquad\qquad\times
\delta(\omega_1+\omega_2-\omega_3-\omega)
\rme^{-\rmi({\bf K}_1+{\bf K}_2-{\bf K}_3)\cdot{\bf v}} .
\end{eqnarray}
If the noncondensate band is taken to be in thermal equilibrium
\begin{eqnarray}\Label{therm-eq}
F({\bf u},{\bf K})&=& {1\over \exp\left( \hbar\omega({\bf K},{\bf u}) -\mu\over 
k_B T\right) -1}  ,
\end{eqnarray}
there is the relation
\begin{eqnarray}\Label{90102}
  G^{(+)}({\bf u},{\bf v}, \omega) &=&
\rme^{(\mu-\hbar\omega)/k_B T}  G^{(-)}({\bf u},{\bf v}, \omega).
\end{eqnarray}
This will still be true even if the equilibrium is merely local, with $ \mu$ 
and $ T$ depending on the position $ {\bf u}$.

Collecting all relevant terms together, we get a master equation term in the 
form
\begin{eqnarray}\fl\Label{s902}
\left. \dot \rho_{C} \right|_{\rm growth} &=&
\int d^3{\bf u}\int d^3{\bf v}
\left[ 
\left\{  G^{(-)}({\bf u},{\bf v},L_{C})
\phi\left({\bf u}-{{\bf v}\over 2}\right)\right\}\rho_{C} ,
\phi^\dagger\left({\bf u}+{{\bf v}\over 2}\right)\right]
\nonumber \\\fl && -
\int d^3{\bf u}\int d^3{\bf v}
\left[ 
\rho_{C} \left\{  G^{(-)}({\bf u},{\bf v},L_{C})
\phi^\dagger\left({\bf u}-{{\bf v}\over 2}\right)\right\},
\phi\left({\bf u}+{{\bf v}\over 2}\right)\right]
\nonumber \\\fl && +
\int d^3{\bf u}\int d^3{\bf v}
\left[ 
\left\{  G^{(+)}({\bf u},{\bf v},L_{C})
\phi^\dagger\left({\bf u}-{{\bf v}\over 2}\right)\right\}\rho_{C} ,
\phi\left({\bf u}+{{\bf v}\over 2}\right)\right]
\nonumber \\\fl && -
\int d^3{\bf u}\int d^3{\bf v}
\left[ 
\rho_{C} \left\{  G^{(+)}({\bf u},{\bf v},L_{C})
\phi\left({\bf u}-{{\bf v}\over 2}\right)\right\},
\phi^\dagger\left({\bf u}+{{\bf v}\over 2}\right)\right] .
\end{eqnarray}

\subsection{Scattering terms}
These come from terms involving two $ \phi$ operators;
of these terms, only those involving one $ \phi$ and one 
$ \phi^\dagger$  can yield a resonant term, corresponding to scattering of a 
non-condensate particle by the condensate. The effect of the non-resonant terms 
is neglected.

Thus we arrive at an approximation to $ H^{(2)}_{I} $ in the form of 
 a term
\begin{eqnarray}\Label{bba1302}
H^{(2)}_{I} &\approx&
2\int d^3{\bf x}\,Z_{2}({\bf x})U({\bf x})
+{\rm h.c.},
\end{eqnarray}
where
\begin{eqnarray}\Label{bba1303}
Z_{2}({\bf x})&=&  u \psi_{NC} ({\bf x})\psi_{NC}^\dagger ({\bf x}),
\\ \Label{st101}
U({\bf x}) &\equiv& \phi^\dagger({\bf x})\phi({\bf x}).
\end{eqnarray}
The term analogous to (\ref{bba8}) in the master equation becomes,
using the notation of (\ref{notationsNC},\ref{notationsC}),
\begin{eqnarray}\Label{bba14}
&& -{4\over\hbar^2}
\int_{-\infty}^{\,0} d\tau\,U({\bf x}) U({\bf x}',\tau)
\int d^3{\bf x}\int d^3{\bf x'}
\langle Z_{2}({\bf x})Z^\dagger_{2}({\bf x'},\tau)\rangle
 \rho_{C}(t).
\end{eqnarray}

\subsubsection{Master equation terms for scattering}
These arise from the term
\begin{eqnarray}\Label{st1}
\int d^3{\bf x}\int d^3{\bf x}'\int_{-\infty}^{\,0} d\tau\,\,\left\langle
\left[Z_2({\bf x})U({\bf x}),\left[Z_2({\bf x}',\tau)U({\bf x}',\tau),
\rho\right]\right]\right\rangle_{\rm NC}.
\end{eqnarray}
We define
\begin{eqnarray}\fl\Label{st2}
M({\bf u},{\bf v},\omega) &=&
{2u^2\over (2\pi)^5\hbar^2}\int d^3{\bf K}_1\int d^3{\bf K}_2\,
F({\bf K}_1,{\bf u})[1+F({\bf K}_2,{\bf u})]
\rme^{\rmi({\bf K}_1-{\bf K}_2)\cdot{\bf v}}
\nonumber\\
\fl&&\qquad\qquad\times 
\delta(\omega_1-\omega_2-\omega),
\end{eqnarray}
and  if $ F({\bf K},{\bf u}) $ corresponds to thermal equilibrium, as in
(\ref{therm-eq}), this  satisfies the relation
\begin{eqnarray}\Label{st3}
M({\bf u},{\bf v},\omega) 
&=& \rme^{-\hbar\omega/k_B T}M({\bf u},{\bf v},-\omega) .
\end{eqnarray}
The terms in the master equation arising from this part become
\begin{eqnarray}\fl\Label{st4}
\left.\dot \rho_{C} \right |_{\rm scatt} &=&
\int d^3{\bf u}\int d^3{\bf v}
\left[ U\left({\bf u}+{{\bf v}\over 2}\right),
\left\{M({\bf u},{\bf v},L_{C})U\left({\bf u}-{{\bf v}\over 2}\right)\right\}
\rho_{C} 
\right]
\nonumber\\\fl&& +
\int d^3{\bf u}\int d^3{\bf v}
\left[\rho_{C} \left\{M({\bf u},{\bf v},-L_{C})U\left({\bf u}-{{\bf v}\over 2}
\right)
\right\},
 U\left({\bf u}+{{\bf v}\over 2}\right)\right].
\end{eqnarray}

\subsection{Terms involving three $ \phi$ operators}\Label{3 phi}
The part of the interaction Hamiltonian can be written
\begin{eqnarray}\Label{bba14010101}
H^{(3)}_{I} &=&
 u\int d^3{\bf x}\,\psi_{NC}({\bf x})
\phi^\dagger({\bf x})\phi^\dagger({\bf x})\phi({\bf x})
+{\rm h.c.}
\end{eqnarray}
The master equation term this time takes the form
\begin{eqnarray}\fl\Label{bba140102}
&& -{u^2\over\hbar^2}
 \int_{-\infty}^{\,0} d\tau\int d^3{\bf x}\int d^3{\bf x'}\,
\phi^\dagger({\bf x}',\tau)\phi({\bf x}',\tau)\phi({\bf x}',\tau)
\phi^\dagger({\bf x})\phi^\dagger({\bf x})\phi({\bf x})
\nonumber \\ \fl && \qquad\qquad\qquad\qquad\qquad\qquad\times
\langle\psi_{NC}({\bf x})\psi_{NC}^\dagger({\bf x'})\rangle
 \rho_{C}(t).
\end{eqnarray}
These terms are probably very small, and will not be included in our analysis.

\subsection{Connection with the \FTGPE\ of Davis \emph{\etal}}
The master equation terms described above are related to the terms in
the \FTGPE, (29a-d) of ~\cite{Davis2001a}.  We have
\begin{eqnarray}\fl\qquad
\mbox{Growth terms} 
&\longleftrightarrow&
U_0\hat{\mathcal{P}}\left\{\langle\hat{\eta}^{\dag}({\bf x})
\hat{\eta}({\bf x})\hat{\eta}({\bf x})\rangle 
\right\},\nonumber\\\fl\qquad
\mbox{Scattering terms} 
&\longleftrightarrow&
U_0\hat{\mathcal{P}}\left\{2\psi({\bf x})\langle\hat{\eta}^{\dag}({\bf x})\hat{
\eta}({\bf x})\rangle \right\}
,\nonumber
\\\fl\qquad
\mbox{Terms with three $\phi$ operators} 
&\longleftrightarrow&
U_0\hat{\mathcal{P}}\left\{2|\psi({\bf x})|^2
\langle \hat{\eta}({\bf x})\rangle
+
\psi({\bf x})^2 \langle\hat{\eta}^{\dag}({\bf x})\rangle \right\}
\label{three}
\nonumber
\end{eqnarray}
We point out that the \FTGPE\ scattering term also includes the forward
scattering part of the Hamiltonian in (\ref{H11}) of this paper.  
The anomalous term of the FTGPE, analyzed in \Sect 5.2 of
\cite{Davis2001a} is mainly responsible for the replacement of the true
interatomic potential by the contact potential in the equations of motion, and
hence the introduction of the momentum cutoff $\Delta$.  In this paper, as 
noted in \Sect\ref{Bands}, we have already
performed this spatial coarse-graining in the Hamiltonian with the cutoff 
$ \Delta$, and so any anomalous terms  will be very small.

\subsection{The full master equation and its stationary solution}
The full master equation can now be written
\begin{eqnarray}\Label{fme1}
 \dot \rho_{C} &=&
\left. \dot \rho_{C} \right |_{\rm Ham}
+\left. \dot \rho_{C} \right |_{\rm growth}
+\left. \dot \rho_{C} \right |_{\rm scatt},
\end{eqnarray}
where the individual terms are given by  (\ref{H1101}), (\ref{s902}) and
(\ref{st4}).
The stationary solution of each of the terms in the master equation
 (\ref{fme1}), and therefore of the master equation itself,   is given by
the grand canonical form
\begin{eqnarray}\Label{s903}
\rho_{\rm s}&\propto& \exp\left(\mu N_{C} - H_{C}\over k_B T\right).
\end{eqnarray}
This follows: 
\Ppt{i}
From the equality of terms in (\ref{s902}) like
\begin{eqnarray}\fl\Label{s904}
\left\{  G^{(-)}({\bf u},{\bf v},L_{C})
\phi\left({\bf u}-{{\bf v}\over 2}\right)\right\}\rho_{\rm s}
&=&
\rho_{\rm s}\left\{  G^{(+)}({\bf u},{\bf v},L_{C})
\phi\left({\bf u}-{{\bf v}\over 2}\right)\right\},
\end{eqnarray}
which can be derived using using
(\ref{90102}), (\ref{s903})
and the commutation relation 
\begin{eqnarray}\Label{st6}
[N_{C},\phi] &=& -\phi ,
\end{eqnarray}
between  the field operator and the condensate 
band number operator 
\begin{eqnarray}\Label{st905}
N_{C} &=&\int d^3{\bf x}\, \phi^\dagger({\bf x})\phi({\bf x)}.
\end{eqnarray}
\Ppt{ii} 
From the equality of terms in (\ref{st4}) like
\begin{eqnarray}\fl\Label{st5}
\left\{  M({\bf u},{\bf v},L_{C})
U\left({\bf u}-{{\bf v}\over 2}\right)\right\}\rho_{\rm s}
&=& 
\rho_{\rm s}\left\{ M({\bf u},{\bf v},-L_{C})
U\left({\bf u}-{{\bf v}\over 2}\right)\right\}.
\end{eqnarray}
which can be derived using (\ref{st3}), (\ref{s903}) and the fact that 
$ [U({\bf x}),N_C]=0$.

\subsection{The projection into the condensate band}
In exactly the same way as noted in \Sect\ref{cold-collision}, the 
projection 
into the condensate band means that the operator $ L_C$ has a projected 
expression when it acts on the condensate band operator $ \phi({\bf x})$
\begin{eqnarray}\fl\Label{s1000}
\rmi{L_C\phi({\bf x})}&=&
{\cal P}_C
\left\{-{\hbar^2\over 2m}\nabla^2\phi({\bf x}) +V({\bf x})
 + u \phi^\dagger({\bf x})\phi({\bf x})\phi({\bf x})\right\},
\end{eqnarray}
which arises directly from the fact that the field operator commutation 
relation is
\begin{eqnarray}\Label{s1001}
\left[\phi({\bf x}),\phi^\dagger({\bf x}')\right] 
&=& {\cal P}_C({\bf x},{\bf x}').
\end{eqnarray}
The projector $  {\cal P}_C({\bf x},{\bf x}') $ will occur frequently in the 
remainder of this paper, as an expression of the fact that the field operator 
$ \phi({\bf x})$ and any approximate representations of it must always be 
expressible in terms of the wavefunctions which span the condensate band.

This projector will play a significant r\^ole in the practical implementation 
of the master equations, and is by no means merely a formal requirement.  
It has two principal effects:
\begin{itemize}
\item [i)]  The nonlocality generates  a spatial smoothing function, and 
because the cutoff is in terms of \emph{energy} rather than momentum, the 
smoothing is stronger where the potential $ V({\bf x})$ is larger.
\item [ii)]  At positions where $ V({\bf x}) > E_R$, all wavefunctions in the 
projector are exponentially small, so the projector is essentially zero.  This 
means that the boundary conditions on any simulational grid are simply that any 
representation of $ \phi({\bf x})$ is zero there.
\end{itemize}
Although the projector can be written down quite easily using (\ref{1001}), 
this expression is not computationally simple---efficient numerical 
implementation is essential for the practicality of any 
simulations.

\section{Approximate forms of the master equation}\Label{Approx}
It is not possible to contemplate a numerical solution of the master equation 
in the form given in (\ref{fme1}), but there are two ways in which 
simplifications can be made, which we will call the ``quantum optical'' master 
equation and the ``high temperature'' master equation.  
\subsection{The ``quantum optical'' master equation for a \BEC}
The method chosen in 
\QKIII\  was to 
expand the field operators in eigenoperators of $ L_C$.  Thus, one wrote
\begin{eqnarray}\Label{ap001}
\phi({\bf x}) & = & \sum_m X_m({\bf x}),
\end{eqnarray}
where the eigenoperator requirement is (see \cite{Gardiner1998a}, eq.\,(28))
\begin{eqnarray}\Label{ap002}
L_CX_m({\bf x}) = \epsilon_m X_m({\bf x}).
\end{eqnarray}
This method has two disadvantages
\begin{itemize}
\item [i)]  We cannot calculate the operators $ X_m({\bf x}) $ exactly, 
although when the Bogoliubov approximation is valid, they can be expressed in 
terms of Bogoliubov amplitudes, as was done 
in \QKIII.
\item [ii)]  The resulting master equation (eq\,(50a--f) of 
\QKIII) can only be derived by making a rotating wave or random phase 
approximation, whose validity can be questioned.  However, this master equation 
is of the Lindblad form (\cite{QN} \Sect 5.2.2), and this is a highly 
desirable, though not absolutely essential, property, since it guarantees that 
solutions are density operators with non-negative eigenvalues.
\end{itemize}
Useful results have been derived using this form of the 
master equation.

\subsection{The ``high temperature'' master equation for a \BEC}
\Label{high-temp}
In this case the essential approximation relies on the fact that 
eigenfrequencies of $ L_{C}$ are small compared to the temperature; precisely
\begin{eqnarray}\Label{ap003}
\hbar |\epsilon_m|/k_BT &\ll 1 .
\end{eqnarray}
This is a condition which is very often met, to an accuracy of at least 10\%, 
at temperatures 
which one would normally find in a degenerate Bose gases with a significant 
non-condensate fraction.  In some sense, the nomenclature 
``high temperature'' is misleading, but in a strict sense, it is a fair 
description of the kind of limit contemplated in (\ref{ap003}).
The master equation that can be derived is not  of the Lindblad form, as 
indeed 
is also the case for the unapproximated master equation (\ref{fme1}).  
However, 
as shown in \cite{Munro1996a}, this probably only means that there are  
transient situations arising after unrealistic (although physically acceptable) 
initial conditions which give rise to a non-positive definite density operator.  
Such transients usually evolve on a time scale more rapid than that used to 
derive the master equation, so they have no physical significance.

We now develop the approximations based on the condition (\ref{ap003}).
The functions $ G^{(\pm)}({\bf u},{\bf v},\omega)$ and 
$ M({\bf u},{\bf v},\omega)$ can then be evaluated approximately for 
 $ \hbar\omega \ll k_B T $, the limit in which the stochastic \GPE\ was 
originally derived \cite{Gardiner2002a}.  Under this 
condition we can use (\ref{st3}) to give
\begin{eqnarray}\Label{ap1}
M({\bf u},{\bf v},\omega) &=&
\left(1-{\hbar\omega\over 2k_B T}\right)M({\bf u},{\bf v},0).
\end{eqnarray}
For $ G^{(\pm)}({\bf u},{\bf v},\omega)$ we can take 
 $ G^{(+)}({\bf u},{\bf v},\omega)\approx G^{(+)}({\bf u},{\bf v},0) $ (Using 
for example the formula (153) of \cite{Gardiner1998a}) and then write using (
\ref{90102})
\begin{eqnarray}\Label{ap2}
 G^{(-)}({\bf u},{\bf v},\omega)\approx \left(1 -{\mu-\hbar\omega\over k_B T}
\right)
 G^{(+)}({\bf u},{\bf v},0).
\end{eqnarray}
These two expressions will be used to derive all that follows in this paper, 
but more accurate expressions could be used, involving higher powers of 
$ \hbar\omega/k_BT$, which would extend the range of the approximation, as 
noted by Stoof \cite{Stoof1999a,Duine2001a,Stoof2001a}.  The highest value of 
$ \epsilon_m$ available is of the order of $ E_R-\mu\approx 2\mu$, so that this 
method would be available for smaller chemical potentials and lower 
temperatures.
\subsubsection{Growth terms}
Both of $ G^{(\pm)}$ are sharply peaked as functions of $ {\bf v}$, so we will 
also approximate 
$ \phi({\bf u}\pm {\bf v}/2) \approx \phi({\bf u})$ in the growth terms,
and get the approximate master equation terms
\begin{eqnarray}
\fl\Label{ap3}
\left. \dot \rho_{C} \right|_{\rm growth} &=&
\left(1-{\mu\over k_B T}\right)
\int d^3{\bf x}\, \bar G ({\bf x})\left\{
\left[\phi({\bf x})\rho_{C} ,\phi^\dagger({\bf x})\right]
 -
\left[ \rho_{C} \phi^\dagger({\bf x}),\phi({\bf x})\right]\right\}
\nonumber \\\fl && +{\hbar\over k_B T}
\int d^3{\bf x}\, \bar G ({\bf x})\left\{
\left[\left(L_C\phi({\bf x})\right)\rho_{C} ,\phi^\dagger({\bf x})\right]
 -
\left[ \rho_{C} \left(L_C\phi^\dagger({\bf x})\right),\phi({\bf x})\right]
\right
\}
\nonumber \\\fl && +
\int d^3{\bf x}\, \bar G ({\bf x})\left\{
\left[ \phi^\dagger({\bf x})\rho_{C} ,\phi({\bf x})\right]
  -
\left[ \rho_{C} \phi({\bf x}),\phi^\dagger({\bf x})\right]\right\}.
\end{eqnarray}
Here we have used the notation
\begin{eqnarray}\Label{ap501}
\bar G({\bf x}) &\equiv & \int d^3{\bf v}\,G^{(+)}({\bf x},{\bf v},0).
\end{eqnarray}

\subsubsection{Scattering terms}
We cannot make the approximation
 $ \phi({\bf u}\pm {\bf v}/2) \approx \phi({\bf u})$ in the scattering terms, 
because 
$  \int d^3{\bf v}\,M({\bf x},{\bf v},0)$ is ill defined, since it involves the 
product $ \delta(\omega_1-\omega_2)\delta({\bf K}_1-{\bf K_2})$.  In this case 
we must keep the full dependence on $ {\bf v}$:
\begin{eqnarray}
\fl\Label{ap4}
\left.\dot \rho_{C} \right |_{\rm scatt} &=&-
\int d^3{\bf x} d^3{\bf v} \, M ({\bf x},{\bf v},0)\bigg\{
\left[ U({\bf x}+{\bf v}/2),[U({\bf x}-{\bf v}/2),\rho_{C} ]\right]
\nonumber\\ \fl &&\qquad\qquad\qquad\qquad
 +{\hbar\over 2k_B T}
\left[U({\bf x}+{\bf v}/2),[\left(L_C U({\bf x}-{\bf v}/2)\right),\rho_{C} ]_+
\right]\bigg
\}.
\nonumber\\\fl
\end{eqnarray}
\subsubsection{Estimate of the amplitude}
To understand the nature of the scattering term, one can evaluate the Fourier 
transform
\begin{eqnarray}\fl\Label{ap401}
\tilde M({\bf x},{\bf k},0) &\equiv & 
{1\over (2\pi)^3}\int d^3{\bf v}\, \rme^{-\rmi{\bf k}\cdot{\bf v}}
M({\bf x},{\bf v},0),
\nonumber \\ \fl &=&
{2u^2\over (2\pi)^5\hbar^2}\int d^3{\bf K}_1\int d^3{\bf K}_2\,
F({\bf K}_1,{\bf x})[1+F({\bf K}_2,{\bf x})]
\delta({\bf K}_1-{\bf K}_2-{\bf k})
\nonumber\\ \fl
\fl&&\qquad\qquad\times
\delta(\omega_1-\omega_2).
\end{eqnarray}
Since $F({\bf K},{\bf x}) $ depends on
\begin{eqnarray}\Label{ap40101}
 e_{\bf K}({\bf x})&\equiv &{\hbar^2 {\bf K}^2\over 2m}+ V({\bf x})
\equiv \hbar\omega({\bf K},{\bf x}),
\end{eqnarray}
this can be written as
\begin{eqnarray}\fl\Label{ap402}
\tilde M({\bf x},{\bf k},0) &=& 
{2u^2\over (2\pi)^5\hbar^2}\int d^3{\bf K}\,
F(e_{\bf K},{\bf x})[1+F(e_{\bf K},{\bf x})]
{2m\over\hbar}\delta(2{\bf K}\cdot{\bf k}- {\bf k}^2).
\end{eqnarray} 
We  can choose the $ x$-axis in the  $ {\bf K}$-integration parallel to 
$ {\bf k}$ and and use the delta function to eliminate $ K_x \to |{\bf k}|/2$.
Using the notation  $ \hat{\bf K}\equiv (K_y,K_z)$, so that
\begin{eqnarray}\Label{ap40104}
 e_{\bf K}({\bf x})&\to  &\hat e_{\hat{\bf K}}({\bf k},{\bf x}) 
\equiv {\hbar^2(\hat{\bf K}^2+|{\bf k}|^2/4)\over 2m}
 +V({\bf x}),
\end{eqnarray}
we can write
\begin{eqnarray}\fl\Label{ap403}
\tilde M({\bf x},{\bf k},0) &= & 
{2mu^2\over (2\pi)^5\hbar^3|{\bf k}|}\int d^2\hat{\bf K}\,
f_{T,\mu}\left(\hat e_{\hat{\bf K}}({\bf k},{\bf x})\right)
\left\{
1+f_{T,\mu}\left(\hat e_{\hat{\bf K}}({\bf k},{\bf x})\right)\right\},
\end{eqnarray}
where
\begin{eqnarray}\Label{ap40301}
f_{T,\mu}(y) &\equiv& 1\over{\rme^{(y-\mu)/k_BT}-1}.
\end{eqnarray}
Using (\ref{ap40104}), we can change the integration variable to 
$ \hat e_{\hat{\bf K}}({\bf k},{\bf x}) $, where the lower limit of 
the integration is now
\begin{eqnarray}\Label{ap40302}
E_{\rm min}({\bf k},{\bf x}) &\equiv & 
\max\left(E_R,V({\bf x})+\hbar^2|{\bf k}|^2/8m\right)/k_BT.
\end{eqnarray}
The integral then can be written
\begin{eqnarray}\fl\Label{ap404}
\tilde M({\bf x},{\bf k},0) &=& 
{2mu^2\over (2\pi)^5\hbar^3|{\bf k}|}{2\pi mk_bT\over\hbar^2}
\int_{(E_{\rm min}-\mu)/k_BT}^\infty {\exp(y)\,dy\over[\exp(y)-1]^2},
\\ \fl \Label{ap405}
&=&{2mu^2\over (2\pi)^5\hbar^3|{\bf k}|}{2\pi mk_bT\over\hbar^2}
{1\over \exp\big(({E_{\rm min}({\bf k},{\bf x})}-\mu)/k_BT\big) -1},
\\ \fl\Label{ap407}
&\approx& {8a^2(k_BT)^2\over 
(2\pi)^2\hbar({E_{\rm min}({\bf k},{\bf x})}-\mu)|{\bf k}|}.
\end{eqnarray}
The principal dependence on $ {\bf k}$ comes from the $ 1/|{\bf k}|$ term, 
which is most significant for small $ |{\bf k}|$. The operators 
$ U({\bf x}\pm{\bf v}/2)$ are restricted to the condensate band, and thus will 
be significant only for $ {\bf x}$ such that $E_R > V({\bf x}) $.  
Thus the major contribution comes from situations such that 
$E_{\rm min}({\bf k},{\bf x})=E_R $, in which case the \emph{only} $ {\bf k}$ 
dependence comes from the $ 1/|{\bf k}|$ term, and we can write
\begin{eqnarray}\Label{ap408}
\tilde M({\bf x},{\bf k},0) &\to & {{\cal M}\over(2\pi)^3|{\bf k}|},
\end{eqnarray}
where we define
\begin{eqnarray}\Label{40801}
{\cal M}&\equiv &  {8a^2(k_BT)^2\over 
(2\pi)^2\hbar({E_{R}}-\mu)} .
\end{eqnarray}

\subsubsection{Approximate local form}
The form (\ref{ap407}) is nonlocal, and this may be an important feature.  
Nevertheless, we can  
try to write an approximate local equivalent in the form (which is essentially 
of the form of the \emph{quantum Brownian motion master equation} as described 
in \cite{QN} \Sect 3.6.1)
\begin{eqnarray}\fl\Label{ap502}
\left.\dot \rho_{C} \right |_{\rm scatt} &=&-
\int d^3{\bf x} \, \bar M ({\bf x})\bigg\{
\left[ U({\bf x}),[U({\bf x}),\rho_{C} ]\right]
 +{\hbar\over 2k_B T}
\left[U({\bf x}),[\left(L_C U({\bf x})\right),\rho_{C} ]_+\right]\bigg\}.
\nonumber\\ \fl
\end{eqnarray}
However, the correct choice of $ \bar M ({\bf x}) $ is not straightforward; a 
possible estimate can be made  by taking $ {\bf k}={\bf k}_1-{\bf k}_2$, where 
$ \hbar{\bf k}_i$ are the momenta of two particles in the condensate band.
We then compute the average
of (\ref{ap407}) over the region $ R_C$ such that $ |{\bf k}_{1,2}|\le 
\sqrt{2m\big(E_R-V({\bf x})\big)}/\hbar$, on the assumption of a 
noninteracting excitation spectrum:
\begin{eqnarray}\Label{ap503}
\left.{1\over |{\bf k}_1-{\bf k}_2|}\right |_{\rm av}
&\equiv &
\int_{R_C} { d^3{\bf k}_1 d^3{\bf k}_2\over |{\bf k}_1-{\bf k}_2|}
\bigg / \int_{R_C} d^3{\bf k}_1 d^3{\bf k}_2,
\\ \Label{ap504}
&=& {6\hbar\,\Theta\big({E_R-V({\bf x})}\big)\over 5\sqrt{2m
\big(E_R-V({\bf x})\big)}}.
\end{eqnarray}
The singularity when $ E_R=V({\bf x})$ is harmless, since the occupation is 
also zero there.  Thus we can choose
\begin{eqnarray}\Label{ap505}
\bar M ({\bf x}) &=& 
{16\pi a^2(k_BT)^2\over \hbar({E_R}-\mu)}
\left.{1\over |{\bf k}_1-{\bf k}_2|}\right |_{\rm av}.
\end{eqnarray}
\subsubsection{Comparison of local and full forms}
There are obvious technical advantages in using the local form 
 (\ref{ap502}) of the scattering term instead of the full form (\ref{ap4}). 
The local form will give the correct stationary distribution for any choice of 
$ \bar M ({\bf x}) $, and is in at least this sense acceptable.  The main 
difference is that the local form gives essentially the same scattering rate 
at all momentum transfers, whereas the full form gives a significant drop off 
at high momentum transfers, and this is known to be a significant feature of 
the kinetics when this is treated by forms of the quantum Boltzmann equation, 
as noted for example in the papers of Svistunov, Kagan and 
Shlyapnikov~\cite{Svistunov1990a,Kagan1992a}, the idea of ``flux'' in energy 
space is used, based on the \emph{locality} in energy space of the collision 
integral.

\section{Wigner function representation}\Label{Wigner}
\subsection{Fokker-Planck equations}\Label{Fokker}
Fokker-Planck equations can be derived by using the transforms of \cite{QN}, as 
described in \cite{Gardiner2002a}.  None of the three principal terms 
(Hamiltonian, growth, scattering) can be put exactly in the form of a genuine 
probabilistic  Fokker-Planck equation, but on the assumption 
that higher order  derivatives 
become less significant, an approximate probabilistic   Fokker-Planck, 
valid for large occupations, can be derived.

The methodology used can be demonstrated for  one part of 
$\left. \rho _{C}\right|_{\rm growth}$:
\begin{eqnarray}\Label{FP1}
\int d^3{\bf x}\,\bar G({\bf x})\left[\left[\phi({\bf x}),\rho_{C} \right] 
-{1\over k_BT}\{\mu\phi({\bf x})- \hbar{\LC}\phi({\bf x}) \}\rho_{C} \, ,
\,\phi^\dagger({\bf x}) \right].
\end{eqnarray}
The transformation to a   Fokker-Planck equation for the Wigner 
function $ W$ of the phase space field function $ \alpha({\bf x})$ is achieved 
by the mappings:
\begin{eqnarray}\Label{FP2}
[\rho_{C} ,\phi^\dagger({\bf x})] &\,\to \,& 
{\deltaC  W\over\deltaC \alpha^*({\bf x})},
\\
\Label{FP3}
[\phi({\bf x}),\rho_{C} ] &\to & {\deltaC  W\over\deltaC \alpha({\bf x})},
\\ \Label{FP4}
\mu\phi({\bf x})\rho_{C}  &\to & \mu\alpha({\bf x}) W,
\\ \Label{FP5}
\hbar{L_C}\phi({\bf x})\rho_{C}  &\,\to  \,&  \LC \alpha({\bf x}) W.
\end{eqnarray}
In this equation there are some important points of notation and approximation
\begin{itemize}
\item [a)] The fact that the condensate band is spanned by the finite set of 
wavefunctions $ Y_n({\bf x})$---see (\ref{1001})---means that we define the 
phase space amplitude  by
\begin{eqnarray}\Label{FP501}
\alpha({\bf x}) &\equiv& \sum_n Y_n({\bf x})\alpha_n ,
\end{eqnarray}
(where $ \alpha_n$ are independent mode amplitudes) and that we define a 
modified functional differentiation operator
\begin{eqnarray}\Label{FP502}
{\deltaC \over\deltaC \alpha({\bf x})} &\equiv&
\sum_n Y^*_n({\bf x}){\partial\over\partial \alpha_n}.
\end{eqnarray}
This would be a genuine functional differentiation operator if the full 
set of modes $ Y_n({\bf x})$ were used instead of only the set within the 
condensate 
band.   Notice that from these definitions
\begin{eqnarray}\Label{FP503}
{\deltaC\alpha({\bf x}) \over\deltaC \alpha({\bf x}')} &\equiv&
{\cal P}_C({\bf x},{\bf x}'),
\end{eqnarray}
whereas the right hand side is a delta function for a true functional 
differentiation operator.
\item[b)]
Including the effect of the mean field arising from the the non-condensate as 
an effective potential
\begin{eqnarray}\Label{FP504}
V_{\rm eff}({\bf x}) &\equiv&
V({\bf x}) + 2u\bar n_{NC}({\bf x}),
\end{eqnarray}
the projected Gross-Pitaevskii operator is introduced as 
\begin{eqnarray}\fl\Label{FP6}
\LC \alpha({\bf x}) & \equiv & {\cal P}_C
\left\{-{\hbar^2\nabla^2\alpha({\bf x})\over 2m}+V_{\rm eff}({\bf x}) 
\alpha({\bf x}) +u| \alpha({\bf x}) |^2 \alpha({\bf x})
\right\}.
\end{eqnarray}
The projector arises as a consequence of (\ref{FP2},\ref{FP3},\ref{FP503}) etc.
\item [c)]
The two replacements (\ref{FP4},\ref{FP5}) are approximate, not only in 
the sense that higher 
order  derivatives are neglected, but also in the sense that 
{\em all}  derivatives which should arise from the left hand side are 
neglected.  This is not an essential 
aspect of the method, since we can technically still handle the first order 
derivative terms which would turn up in the right hand sides of 
(\ref{FP4},\ref{FP5}), and these would produce only second order derivatives 
in the final  Fokker-Planck equation.
This would  produce a fractional correction of order $ \mu/k_BT$ to 
the noise terms only.
\end{itemize}
The projected Gross-Pitaevskii operator can be written in terms of the
Gross-Pitaevskii Hamiltonian 
\begin{eqnarray}\Label{FP8}
H_{\rm GP}&\equiv & \int d^3{\bf x}\,\left\{
{\hbar^2|\nabla\alpha({\bf x})|^2\over 2m}
+V_{\rm eff}({\bf x}) |\alpha({\bf x})|^2
+{u\over 2}| \alpha({\bf x}) |^4\right\},
\end{eqnarray}
in the form
\begin{eqnarray}\Label{FP601} 
\Label{FP7}
\LC \alpha({\bf x})
&=&{\deltaC  H_{\rm GP}\over\deltaC \alpha^*({\bf x})}.
\end{eqnarray}

We also define the Gross-Pitaevskii particle number
\begin{eqnarray}\Label{FP801}
N_{\rm GP} &\equiv & \int d^3{\bf x}\,|\alpha({\bf x})|^2.
\end{eqnarray}
The term (\ref{FP1}) then transforms to the term in the   
Fokker-Planck equation
\begin{eqnarray}\Label{FP9}
\int d^3{\bf x}\,\bar G({\bf x}){\deltaC \over\deltaC \alpha({\bf x})}\left
\{{\deltaC  W\over\deltaC \alpha^*({\bf x})}
-{W\over k_B T}{\deltaC  
(\mu N_{\rm GP} - H_{\rm GP})\over\deltaC \alpha^*({\bf x})}
\right\}.
\end{eqnarray}
Using the same kind of approximations, the scattering term  
(\ref{ap502}) can be transformed to a corresponding form, and we obtain, 
putting all terms together,  the   Fokker-Planck equation
\begin{eqnarray}\fl\Label{FP10}
{\partial W\over\partial t}&=&\int d^3{\bf x}\,
{\deltaC \over\deltaC \alpha({\bf x})}\bigg\{
{\rmi W\over\hbar}{\deltaC   H_{\rm GP}\over\deltaC \alpha^*({\bf x})}
 + \bar G({\bf x})\left[
{\deltaC  W\over\deltaC \alpha^*({\bf x})}-{W\over k_B T}{\deltaC  
(\mu N_{\rm GP} - H_{\rm GP})
\over\deltaC \alpha^*({\bf x})}\right]\bigg\} 
\nonumber\\\fl
&&-\int d^3{\bf x}\,
{\deltaC \over\deltaC \alpha^*({\bf x})}\bigg\{
{\rmi W\over\hbar}{\deltaC   H_{\rm GP}\over\deltaC \alpha({\bf x})}
 - \bar G({\bf x})\left[
{\deltaC  W\over\deltaC \alpha({\bf x})}-{W\over k_B T}{\deltaC  
(\mu N_{\rm GP} - H_{\rm GP})
\over\deltaC \alpha({\bf x})} \right]\bigg\} 
\nonumber\\ \fl &&
+\int d^3{\bf x}\,d^3{\bf x}'\,
M\left({{\bf x}+{\bf x}'\over 2},{\bf x}-{\bf x}',0\right)
\left[{ \alpha^*({\bf x}){\deltaC \over\deltaC  \alpha^*({\bf x})} 
-\alpha({\bf x}){\deltaC \over\deltaC  \alpha({\bf x})}}
\right]
\nonumber\\\fl &&\times\Bigg\{
 \alpha^*({\bf x}')\left[{\deltaC  W\over\deltaC  \alpha^*({\bf x}')} 
+
{W\over k_BT}
{\deltaC  H_{\rm GP}\over\deltaC  \alpha^*({\bf x}')}\right]
 -\alpha({\bf x}')\left[{\deltaC  W\over\deltaC  \alpha({\bf x}')}
+
{W\over k_BT}
{\deltaC  H_{\rm GP}\over\deltaC  \alpha({\bf x}')}\right]
\Bigg\}.
\nonumber \\ \fl
\end{eqnarray}

\subsection{Stationary solutions of the  Fokker-Planck 
equation}
Independently of the forms of $ \bar G({\bf x})$ and $ M({\bf x},{\bf x}')$,
the stationary solution of the   Fokker-Planck (\ref{FP10})
equation is given by
\begin{eqnarray}\Label{FP11}
W_{\rm s}&\propto & \exp\left({\mu N_{\rm GP}-H_{\rm GP}\over k_BT}\right).
\end{eqnarray}
 This is the grand canonical distribution expected for a classical field theory
 whose field $\alpha({\bf x},t)$ obeys the \GPE.

\subsection{Stochastic \GPE s}
The   Fokker-Planck equation (\ref{FP10}) is equivalent to a set of 
\SDE s, which we shall write in the \emph{Stratonovich} form, in a choice of 
forms in which various degrees of simplification are made of the scattering 
terms, as follows. The notation $\mathbf{(S)}$ to the right of an equation
indicates that it is in Stratonovich form as in \cite{SM}.
\subsubsection{Full form of the \SDE s}
Using the full form of the scattering term, we have
\begin{eqnarray}\fl\Label{ap6}
{\bf (S)}\,\,d\alpha({\bf x},t) 
&=& -{\rmi\over \hbar}\LC \alpha({\bf x},t)
\,dt
\nonumber\\\fl&&
+ {\cal P}_C\Bigg\{
{\bar G ({\bf x})\over k_B T}
\left\{\mu \alpha({\bf x},t) -\LC \alpha({\bf x},t)\right\}\,dt
+ dW_{G}({\bf x},t)
\nonumber\\\fl&&
- {1\over 2k_B T}\int d^3{\bf x}'
 M \left({{\bf x}+{\bf x}'\over2},{\bf x}-{\bf x}',0\right)
\alpha({\bf x})
\nonumber\\ \fl &&\times
\left\{\alpha^*({\bf x}',t)\left(\LC \alpha({\bf x}',t)\right)
- \alpha({\bf x}',t)\left(\LC \alpha({\bf x}',t)\right)^*
\right\}
\,dt
\nonumber\\ \fl
&&   \qquad\qquad\qquad\qquad\qquad\qquad\qquad 
+\rmi\alpha({\bf x})\,dW_{M}({\bf x},t)
\Bigg\}.
\end{eqnarray}
The noise  $ dW_G({\bf x},t) $ is complex, while $ dW_M({\bf x},t)$ is real; 
they are independent of each other, and satisfy the relations
\begin{eqnarray}\Label{ap7}
dW^*_G({\bf x},t)dW_G({\bf x}',t) &=&
    2 \bar G ({\bf x}) \delta({\bf x}-{\bf x}')\,dt,
\\ \Label{ap8}
dW_G({\bf x},t)dW_G({\bf x}',t)&=& dW^*_G({\bf x},t)dW^*_G({\bf x}',t)=0,
\\ \Label{ap9}
dW_M\left({\bf x},t\right)dW_M\left({\bf x}',t
\right) &=& 2M \left({{\bf x}+{\bf x}'\over2},{\bf x}-{\bf x}',0\right)
\,dt.
\end{eqnarray}
\subsubsection{Simplified non-local form}
The implementation of the last  two lines of (\ref{ap6}) obviously presents 
some technical issues. However, if we simplify  (\ref{ap407}) by setting 
$ E_{\rm min}({\bf k},{\bf x})\to E_R$ (as discussed there), the nonlocal form 
can be considerably simplified.  The operator $1/\sqrt{-\nabla^2} $, of which 
$ 1/|{\bf k}|$ is the Fourier transform, is not singular when acting on 
functions with well behaved Fourier transform as $ {\bf k}\to 0$.  Using this, 
the \SDE\ becomes
\begin{eqnarray}\fl\Label{ap6s}
{\bf (S)}\,\,d\alpha({\bf x},t) &=& 
-{\rmi\over \hbar}\LC \alpha({\bf x},t)\,dt
\nonumber\\\fl&&
+{\cal P}_C\Bigg\{{\bar G ({\bf x})\over k_B T}
\left\{\mu \alpha({\bf x},t) -\hbar{\LC}{\alpha({\bf x},t)}\right\}\,dt
+ dW_{G}({\bf x},t)
\nonumber\\\fl&&
- {{\cal M\alpha({\bf x})}\over 2k_B T}
{1\over\sqrt{-\nabla^2}}
\left\{\alpha^*({\bf x},t)\LC \alpha({\bf x},t)
- \alpha({\bf x},t)\left(\LC \alpha({\bf x},t)\right)^*
\right\}\,dt
\nonumber\\ \fl && 
 \qquad\qquad\qquad\qquad\qquad\qquad\qquad
+\rmi\alpha({\bf x})\,dW_{\cal M}({\bf x},t)\Bigg\}.
\end{eqnarray}
The noise  $ dW_G({\bf x},t) $ is complex, while $ dW_{\cal M}({\bf x},t)$ is 
real; they are independent of each other, and satisfy the relations
\begin{eqnarray}\Label{ap7s}
dW^*_G({\bf x},t)dW_G({\bf x}',t) &=&
    2 \bar G ({\bf x}) \delta({\bf x}-{\bf x}')\,dt,
\\ \Label{ap8s}
dW_G({\bf x},t)dW_G({\bf x}',t)&=& dW^*_G({\bf x},t)dW^*_G({\bf x}',t)=0,
\\ \Label{ap9s}
dW_{\cal M}({\bf x},t)dW_{\cal M}({\bf x}',t) 
&=& {2{\cal M}\over\sqrt{-\nabla^2}} \delta({\bf x}-{\bf x}')\,dt.
\end{eqnarray}
\subsubsection{Local form of the \SDE s}
These take the form
\begin{eqnarray}\fl\Label{ap6fl}
{\bf (S)}\,\,d\alpha({\bf x},t) &=&
-{\rmi\over \hbar}\LC \big(\alpha({\bf x},t)\big)\,dt
\nonumber\\\fl&&
+{\cal P}_C\Bigg\{{\bar G ({\bf x})\over k_B T}
\left\{\mu \alpha({\bf x},t) -\hbar{\LC}{\alpha({\bf x},t)}\right\}\,dt
+ dW_{G}({\bf x},t)
\nonumber\\\fl&&
- {\bar M ({\bf x})\alpha({\bf x})\over 2k_B T}
\left\{\alpha^*({\bf x},t)\LC \alpha({\bf x},t)
- \alpha({\bf x},t)\left(\LC \alpha({\bf x},t)\right)^*
\right\}
\,dt
\nonumber\\\fl&&
 \qquad\qquad\qquad\qquad\qquad\qquad\qquad
 + \rmi\alpha({\bf x})\,dW_{\bar M}({\bf x},t)\Bigg\}.
\end{eqnarray}
The noise  $ dW_G({\bf x},t) $ is complex, while $ dW_{\bar M}({\bf x},t)$ is 
real; 
they are independent of each other, and satisfy the relations
\begin{eqnarray}\Label{ap7fl}
dW^*_G({\bf x},t)\,dW_G({\bf x}',t) &=&
    2 \bar G ({\bf x}) \,\delta({\bf x}-{\bf x}')\,dt,
\\ \Label{ap8fl}
dW_G({\bf x},t)\,dW_G({\bf x}',t)&=& dW^*_G({\bf x},t)\,dW^*_G({\bf x}',t)=0,
\\ \Label{ap9fl}
dW_{\bar M}({\bf x},t)\,dW_{\bar M}({\bf x}',t) 
&=& 2 \bar M({\bf x})\,\delta({\bf x}-{\bf 
x}')
\,dt.
\end{eqnarray}
\subsection{Comparison with other methods}\Label{Stoof}
The \SDE s we have derived have  similarities with those we have previously 
derived, as well as with Stoof's.  However,
our way of implementing the idea of eliminating higher energy thermalized 
modes has significant differences from Stoof's:

\Ppt{i} There are major technical differences in how the elimination
is done.  Ours is based on the ideas of quantum optics, which are used
to develop a quantum mechanical master equation.  The master equation is then
transformed using the Wigner function to give a \FPE, which is
equivalent to a \SDE. Stoof uses a functional integral formulation of
the Keldysh method, in which the elimination is achieved in the action
integral.  This method is almost certainly equivalent to our quantum
optical method.  Thus, although the technical methods used appear very
different, this is not physically significant.

\Ppt{ii} However the choice of what to eliminate is different.  Stoof 
eliminates modes with energy   $ \grapprox \mu$, and  finds a 
\FPE\ (equivalent to a \SDE) which involves 
self energy functions, which he evaluates using the many-body T-matrix method.
In contrast we carry out the elimination as a two-stage process, as detailed in 
\Sects{\ref{Desc},\ref{Master}}.  Thus we separate the elimination process 
which is  required to give an effective field theory from that required to give 
a 
master equation for the condensate band.  This has the advantage that we do not 
need to use a many body T-matrix description. 

\Ppt{iii}  The equations which arise are given above as 
(\ref{ap6},\ref{ap6s},\ref{ap6fl}), depending on degrees of approximation 
used.  The principal differences are the appearance of the projector, and the 
inclusion of scattering terms.  There is also the implicit difference that the 
field variable we use is not exactly the same as Stoof's, since it includes a 
wider range of states.  

\Ppt{iv} Our earlier work \cite{Gardiner2002a} included the 
scattering terms, but was unable to give any precise value to them because of 
the crudeness of the methods used, and also for reasons related to the fact, 
which we have found in this paper, that the scattering cannot be described 
locally without making very crude approximations, such as those which lead to 
the form (\ref{ap6fl}).  The explicit appearance of the projector in the 
equations did not occur there or in Stoof's method.  It is possible that in 
numerical implementation, the projector will not always be essential, but we 
are confident that there are situations in which its inclusion will be 
important.

\section{Conclusions and outlook}\Label{conclusions}
The problem we have set ourselves in this paper is to produce a description of 
\emph{finite temperature} condensed or nearly condensed Bose gases which is 
both accurate and implementable.  Apart from those used to derive the basic 
Markovian master equation (\ref{met}), there are only two significant 
approximations 
made in our derivation:
\begin{itemize}
\item [i)] The energy eigenvalues of the excitations are small compared to 
temperature, as noted in \Sect\ref{high-temp}.  This is needed to get a 
master equation of a reasonably simple kind, the ``high temperature master 
equation.''

\item[ii)] The occupations of the modes being treated in the condensate band 
are significantly larger than unity, as noted in \Sect\ref{Fokker}.  This is 
essential to be able to use a Wigner function representation of the master 
equation which becomes equivalent to a \emph{classical field} representation.
\end{itemize}
These two conditions are essentially the same, since the occupation of a mode 
of energy $ \epsilon$ is large if $ \epsilon \ll k_BT$, even though the reasons 
for the conditions are quite independent.
\mjd{Therefore, it is clear that if one is using the ``high temperature 
master equation,'' then it is also sensible to use the Wigner function 
classical field representation, since the two are accurate to the same degree 
of approximation.}
\subsection{The validity of the classical field representation}
Classical field representations have often been introduced heuristically by 
the argument that the quantum field operator can be replaced by a classic field 
function provided the occupations of the modes are high.  The Wigner function 
methodology which we use gives systematic way of implementing \mjd{these}
heuristic  ideas, and gives a way a assessing the validity of the formalism%
\footnote{The alternative P- and Q-function descriptions are better called 
\emph{phase space representations}, since the resulting equations of motion 
possess non-negligible noise terms of a purely quantum nature, and 
thus the field cannot be said to be a \emph{classical field}.}.
It is important to realize that, when formulated through the Wigner function, 
the classical field method gives a truly quantum mechanical description of the 
system, subject only to the technical approximations made.  This means that we 
expect the major quantum mechanical aspects to be correctly treated; in other 
words, the classical field method, correctly and carefully formulated, 
\mjd{has} no 
heuristic \mjd{approximations} introduced to provide refuge from quantum 
mechanics---it is a valid way of treating quantum mechanics in a certain degree 
of approximation, which fortunately behaves very \mjd{much} like a classical 
description.

As a result, certain quantum properties make their presence felt when
using this method. The Wigner function exhibits ``vacuum occupation''
of the modes---the mean square of the classical field is non-zero even
when there are no particles. This follows from the relationship
between the classical field averages and the quantum averages in the
form
\begin{eqnarray}\Label{conc1}
\left\langle a^\dagger a + a a^\dagger\over 2\right\rangle
&=& \int d^2\alpha\,|\alpha |^2W(\alpha,\alpha^*),
\end{eqnarray}
which says that the minimum occupation exhibited by the classical
field is the ``vacuum occupation,'' half a particle per mode---an
expression of the Heisenberg uncertainty principle. When summed over
the infinite number of modes which constitute the full system, this
gives a divergent field function; thus a cutoff must be introduced,
such as the one we have chosen, which defines the boundary of the
condensate band.

There are two different ways in which the classical field method applied to the 
condensate band can be valid.
\begin{itemize}
\item [i)] The temperature may be so low that there is little occupation at the 
top of the condensate band, so that all of the noise terms derived in 
\Sect\ref{Wigner} are zero. At low energies, the excitations of a Bose 
condensate are largely phonon-like, and amount to quantized shape and density 
oscillations of the condensate itself.  These are only noticeable if their 
occupation is large; modes with low occupation, although badly described by the 
classical field method, provide a completely negligible contribution to the 
system.

However, since there is no external noise, in this case the initial occupations 
of the modes must be chosen so as to represent the existence of a finite 
temperature.  Even if the temperature is zero, these occupations must be chosen 
to give the correct ``vacuum occupation,'' and this gives rise to 
irreversible effects, as first noted by Steel \etal\ \cite{Steel1998b}.
\item[ii)]  If the temperature is not very low, the total number of particle 
in modes with occupations $ \ltapprox 1$ can be very significant, and 
this is necessarily the case during the process of condensate growth from a 
vapour. The influence of these modes has to be included, and this is what has 
been done in this paper.

In principle the initial conditions should be chosen as in i) to give 
the right ``vacuum occupation'', but their effect rapidly dies out 
because of the irreversible coupling to the reservoir which generates 
the noise and damping terms.
\end{itemize}

\subsection{The degenerate non-condensed Bose gas}
It is known that during the process of condensate growth the occupations of a 
large number of modes become significantly larger than one before the 
appearance of a single dominant mode, the condensate.  These modes are of 
course describable by our stochastic classical field equations 
(\ref{ap6}--\ref{ap9fl}), since we have nowhere assumed the existence of a 
condensate, and because their high occupation makes the classical field method 
valid.  Kagan, Svistunov and Shlyapnikov
\cite{Kagan1992a} in their pioneering work on the initiation of \BECn, also 
argued that a description in terms of the \GPE\ was possible for such a system.
However, because their philosophy was more qualitative, the technical 
details which we are compelled to attend to do not appear in their 
work.

\subsection{The projector into the condensate band}
One of the essential features of the description is the presence of a
projector in the equations of motion for the classical field. This
prevents non-condensate band modes being wrongly included in numerical
simulations. Such a projector has already been implemented in
\cite{Davis2001a,Davis2001b,Davis2002a}, but only for a homogeneous
system. While the projector for a system with a trap potential can be
written down explicitly as in (\ref{1001}-\ref{1002}), it is not an
easy task to implement this projector efficiently.

The projector deals correctly with two problems which arise in practical 
simulations:
\begin{itemize}
\item [i)] \emph{The fineness of the mesh: } The distance between mesh points 
determines the highest momentum which can be represented in the simulation.  
However, one cannot choose this to be the physical cutoff, even in the case of 
no trapping potential, since nonlinear terms such as
$ {\cal P}_C \left\{|\alpha({\bf x})|^2\alpha({\bf x})\right\}$ in 
(\ref{FP6}) are wrongly represented this way.  To represent this term correctly 
without aliasing, the numerical wavenumber cutoff must \mjd{extend to at least 
\emph{two} times the physical cutoff.}
If one assumes the physical cutoff is to 
be given by the mesh fineness, \mjd{incorrect}
 results will be obtained unless there is 
no significant occupation above one \mjd{half} of the mesh wavenumber cutoff.

In addition, when a trap potential is present, a cut at a minimum wavelength 
does not correspond to a cut at definite energy.  This would mean that even 
non-interacting atoms would pass from the condensate band to the non-condensate 
band with the progression of time, presenting some difficulties for the 
formalism

\item [ii)] \emph{The boundary condition at the edge of the spatial grid: }
Because the projector involves trap eigenfunctions only up to energy $ E_R$, 
in the case of a non-vanishing trap potential
there is a distance $ R $ such that for $ |{\bf x}| > R $ projected functions 
become exponentially small.  This means that not only are the field functions
$ \alpha({\bf x})$ of \Sect\ref{Wigner} exponentially small there, but also 
the added noise terms as well.  
Therefore in practical simulations on a grid of finite size, the issue of 
the boundary condition for the noise and the field function at the edge of the 
grid is determined; the simulation region must be so large that all field and 
noise functions can be set equal to zero at the boundary.

Where there is no trap potential, the box inside which the simulation is being 
implemented has real physical significance, and periodic boundary conditions 
are usually chosen.  
\end{itemize}

\subsection{Applications}
\begin{itemize}
\item [i)] \emph{Condensate growth: } The theory of condensate growth
is at this stage not entirely satisfactory---there is still
disagreement between theory and experiment under certain conditions
\cite{Davis2002b,Koehl2002a}. The main defect in computations of
condensate growth
\cite{Davis2002b,Koehl2002a,Gardiner1997b,Gardiner1998b,Lee2000a,%
Davis2000a,Bijlsma2000a} to date is an inadequate treatment of the phonon-like 
quasiparticle excitations.  Using the formulation proposed here would 
definitely include these correctly.

\item [ii)] \emph{Vortex lattice growth: } The theoretical situation
in the growth of vortex lattices is in a very preliminary stage. The
stabilization of the vortices into a lattice is clearly a result of
some kind of irreversible process, for which a phenomenological model
was first proposed by Tsubota \etal~\cite{Tsubota2002a}, while we
presented a physically based model~\cite{Penckwitt2002a} based on a
rotating frame version of our phenomenological growth
equation~\cite{Gardiner2002a} which attributed the necessary
irreversibility to interaction with a thermal cloud.

In contrast, Lobo \etal~\cite{Lobo2003a} used a simple stochastic
classical field model of the type mentioned above, in which the noise
arises from initial conditions with no interaction with a thermal
cloud to provide the requisite damping. However, the noise in their
implementation of this model is nonzero throughout the simulation
region, and this causes some difficulty in deciding the appropriate
boundary condition at the edge of the simulation region, since they
choose the high energy cutoff to be given by the fineness of the
simulation mesh, with no projector as in this paper. In effect, this
means that the vacuum can transfer angular momentum to the system, and
that indeed the vacuum is different depending on whether the
simulation is carried out in a rotating frame or a non-rotating frame.

Our proposed methodology would avoid the boundary condition
difficulties of the last model, combining its ideas with those of our
earlier model.

\item [ii)] \emph{Heating of a condensate by mechanical disturbance: }
When a condensate is mechanically disturbed, some heating occurs.  This is 
clearly also a phenomenon which can be treated by our methodology.
\end{itemize} 
\subsection{Technical aspects: }
The proposed stochastic differential equations
(\ref{ap6}--\ref{ap9fl}) are not as simple as one would like, but are
nevertheless probably quite practical even in three dimensions. The
noise and damping arise due to the \emph{growth} and \emph{scattering}
terms in the master equation. The former describe the transfer of
particles between the condensate and non-condensate bands, while the
latter represents the effect of non-condensate particles colliding
with condensate band atoms with one particle remaining in each band.
The growth terms in the stochastic \GPE\ are relatively easy to
implement numerically, but there are definitely some challenges for
the scattering terms. The essential feature of the projector into the
condensate band in the equations is an additional challenge,
relatively easy to conceive, but formidable to implement. However, the
projector is not of merely cosmetic significance, and this challenge
must be faced.

\subsection{Outlook}
The practical implementation of our methodology will be the feature of
forthcoming papers which will include treatments of condensate growth,
vortex lattices and heating.

\ack This work was supported by the Marsden Fund 
of the Royal Society of New Zealand under contract PVT-202, and by the 
Australian Research Council Centre of Excellence grant CE0348178.

\section*{References}
\bibliographystyle{prsty}
\bibliography{SGPEII}
\end{document}